\documentclass[twocolumn,showpacs,preprintnumbers,floatfix,letterpaper,prc]{revtex4}

\usepackage{citesort,enumerate}
\usepackage{amsmath,amssymb} % AMS math & symbols
\usepackage{bm}              % bold math
\usepackage{amscd}           % for extensible arrows (e.g., limits)
\usepackage{graphicx}        % PostScript figures
\usepackage{hhline,multirow} % for nicer tables
\usepackage{dcolumn}         % Align table columns on decimal point

\newcommand{\eqeqref}[1]{Eq.~\eqref{#1}}
\newcommand{\eqseqref}[1]{Eqs.~\eqref{#1}}
\newcommand{\refref}[1]{Ref.~\cite{#1}}

\newcommand{\beq}{\begin{equation}}
\newcommand{\eeq}{\end{equation}}
\newcommand{\bea}{\begin{eqnarray}}
\newcommand{\beas}{\begin{eqnarray*}}
\newcommand{\beau}[1]{\begin{equation} \label{#1} \begin{array}{rcl}}
\newcommand{\eea}{\end{eqnarray}}
\newcommand{\eeas}{\end{eqnarray*}}
\newcommand{\eeau}{\end{array} \end{equation}}
\newcommand{\bay}{\begin{array}}
\newcommand{\eay}{\end{array}}
\newcommand{\bals}{\begin{align*}}
\newcommand{\eals}{\end{align*}}

\newcommand{\ds}{\displaystyle}

\newcommand{\ra}{{\rightarrow}}

\newcommand{\vev}[1]{\langle #1 \rangle}

\begin{document}

%%%%%%%%%%%%%%%%%%%%%%%%%%%%%%%%%%%%%%%%%%%%%%%%%%%%%%%%%%%%%%%%%%%%%%
%%%%%%%%%%%%%%%%%%%%%%%   TITLE    %%%%%%%%%%%%%%%%%%%%%%%%%%%%%%%%%%%
%%%%%%%%%%%%%%%%%%%%%%%%%%%%%%%%%%%%%%%%%%%%%%%%%%%%%%%%%%%%%%%%%%%%%%

%\preprint{preprint/number}

\title{Formation time scaling and hadronization in cold nuclear
  matter} 

\author{Alberto Accardi}%
  %\email{aaccardi@nt3.phys.columbia.edu}
  %\altaffiliation[Also at ]{Physics Department, XYZ University.}
  %\homepage{http://www.Second.institution.edu/~Charlie.Author}
\affiliation{
Department of Physics and Astronomy, Iowa State University, \\
Ames, Iowa 50011-3160, U.S.A.
}
\date{April 17, 1006}

\begin{abstract}
%Hadron suppression in Deep Inelastic Scattering on nuclei (nDIS) is an
%important tool to study the space-time evolution of the hadronization
%process. Two theoretical frameworks are presently competing to explain the
%observed hadron suppression: energy loss models, with hadron formation
%outside the nuclear medium, and nuclear absorption models, with
%hadronization starting inside the medium. Establishing the leading
%mechanism in nDIS is very important for a correct interpretation of 
%jet quenching in heavy-ion collisions as a signature of the production
%of a Quark-Gluon Plasma.
I propose a scaling analysis of the hadron
multiplicity ratio measured in Deep Inelastic Scattering on nuclear
targets as a tool to distinguish energy loss
and nuclear absorption effects on hadron suppression in cold nuclear matter.
The proposed scaling variable is a function of the hadron fractional
energy and of the virtual photon energy. Its functional form,
which depends on a parameter 
$\lambda$, can be fixed by general theoretical considerations and
encompasses both energy loss and absorption models. 
The parameter $\lambda$ is fitted to HERMES experimental data and
shown to favor prehadron nuclear absorption as leading mechanism for hadron
suppression as opposed to quark energy loss. 
\end{abstract}



\maketitle

%%%%%%%%%%%%%%%%%%%%%%%%%%%%%%%%%%%%%%%%%%%%%%%%%%%%%%%%%%%%%%%%%%%%%%
%%%%%%%%%%%%%%%%%%%%%%%   PAPER    %%%%%%%%%%%%%%%%%%%%%%%%%%%%%%%%%%%
%%%%%%%%%%%%%%%%%%%%%%%%%%%%%%%%%%%%%%%%%%%%%%%%%%%%%%%%%%%%%%%%%%%%%%

\section{Introduction}
\label{sec:intro}

In Deep Inelastic Scattering on nuclear targets (nDIS) 
one observes a suppression of hadron production
\cite{EMC,HERMES1,HERMES2,HERMES3,vanderNat,CLAS,HERMES2H} 
analogous to hadron quenching in heavy-ion collision at the
Relativistic Heavy-Ion Collider (RHIC) \cite{RHIC}.

The cleanest environment to address nuclear modifications of hadron
production is nuclear DIS: it allows to experimentally control
many kinematic variables; the nuclear medium, i.e., the nucleus
itself, is well known; the multiplicity in the final state is low. 
Moreover, the nucleons act as femtometer-scale detectors 
allowing to experimentally study the propagation of a parton 
in this ``cold nuclear matter'', and its space-time evolution into the
observed hadron. In the case of heavy ion collisions,  
one wants to use hadron suppression as a tool to
extract the properties of the hot and dense system created in the
collision, also called ``hot nuclear matter''. 
%To this purpose we need
%to develop well calibrated computational tools to
%relate the magnitude of hadron suppression to properties of the hot 
%medium like its density and temperature. 
If, for example, the parton's color were neutralized on much larger
scales than the nuclear radius, hadron suppression would be
attributed to parton energy loss \cite{jettomreview}. 
Analysis of midrapidity hadron production at RHIC in the energy loss
framework leads to a
medium temperature $T \approx 400$ MeV, well in excess of the critical
temperature $T_c \approx 170$ MeV for the transition into a deconfined
Quark-Gluon Plasma \cite{VitevQM05,Wang}.
If, on the contrary, color
neutralization started on the nuclear radius scale or before, one
should also account for the interactions of the medium with the
prehadron, the color neutral precursor of the hadron
\cite{Cassing:2003sb}. This would lead to a different, presumably 
lower, value of the medium temperature.
Knowing precisely how the struck quark propagates in cold nuclear matter
-- most importantly, whether it starts hadronizing inside or outside
the nuclear medium -- is essential for correctly using
hadron quenching as a signature of the production of a Quark-Gluon
Plasma at RHIC. 

Experimental data on hadron production in nDIS are usually presented
in terms of the multiplicity ratio
\cite{EMC,HERMES1,HERMES2,HERMES3,vanderNat,CLAS} 
\begin{align}
  R_M^h(z_h,\nu) = \frac{1}{N_A^{DIS}}\frac{dN_A^h}{dz_hd\nu} \Bigg{/}
    \frac{1}{N_D^{DIS}}\frac{dN_D^h}{dz_h d\nu} ,\
    \label{MultiplicityRatio}	   
\end{align}
i.e., the single hadron multiplicity per DIS event 
on a target of mass number $A$ 
normalized to the multiplicity on a deuterium target, as a
function of the virtual photon energy $\nu$ and of $z_h=p\cdot p_h /
p\cdot q$, with $p$ the target 4-momentum divided by A, 
$p_h$ the hadron 4-momentum
and $q$ the virtual photon 4-momentum. In the target rest frame
$z_h=E_h/\nu$ is the hadron fractional energy with
respect to the virtual photon energy.
The double ratios in \eqref{MultiplicityRatio} cancel to a large extent
initial state effects like the modifications of parton distribution
functions due to shadowing and EMC effects, exposing the nuclear
modifications of the fragmentation process. If no nuclear effects
modified the fragmentation process, we would expect $R_M \approx 1$. 
In fact, what is experimentally observed
\cite{EMC,HERMES1,HERMES2,HERMES3} is a suppression of 
pions, kaons and antiprotons in the $z_h=0.1-1$ and $\nu=7-100$ GeV
range. Protons are enhanced at $z_h\lesssim 0.4$ (``proton anomaly'')
and suppressed above. Both quenching and enhancement increase with $A$.  

Despite a lot of experimental and theoretical efforts, 
the leading physical mechanism for hadron quenching in nDIS
has not yet been unambiguously established. In particular, as shown in
\cite{Accardi05,AGMP05}, the observed approximate 
$A^{2/3}$ scaling of the experimental 
data cannot distinguish models based on nuclear absorption
\cite{Accardi05,AMP03,AGMP05,Kopeliovich,Falteretal04,BG87} from models
based on parton energy loss \cite{Wang,Arleo}, as is often assumed.
Indeed, single hadron suppression in nDIS
obeys a $A^{2/3}$ law (broken at $A \gtrsim 80$) in both energy loss
and absorption models  \cite{Accardi05}.  Even the more refined analysis 
in terms of $R_M=cA^\alpha$ fits proposed in \cite{AGMP05}
cannot clearly distinguish the 2 classes of models.

In this paper, I propose a scaling analysis of $R_M$ as a tool to
disentangle parton energy loss and nuclear absorption effects on hadron
production in nDIS. More in detail, I conjecture that $R_M$ should not
depend on $z_h$ and $\nu$ separately but should
depend on a combination of them:
\begin{align}
  R_M = R_M\big[\tau(z_h,\nu)\big] \ ,
 \label{eq:RMscaling}
\end{align}
where the scaling variable $\tau$ is defined as
\begin{align}
  \tau & = C\, z_h^\lambda (1-z_h) \nu \ .
 \label{eq:scalingvar}
\end{align}
The scaling exponent $\lambda$ is introduced as a way of
approximating and summarizing the scaling behavior of experimental
data and theoretical models. It will be separately obtained by a best fit
analysis of data and theoretical computations, see
Section~\ref{sec:fitprocedure}.    
The proportionality constant $C$ cannot
be determined by the fit. 
A possible scaling of $R_M$ with $Q^2$ is not considered in this
analysis because of its model dependence; moreover, in the HERMES data
considered in this paper, the dependence of the average $\vev{Q^2}$ on
$z_h$ and $\nu$ is very mild, 
implying very small effects on the scaling of $R_M$.

As discussed in Section~\ref{sec:scalingtheory}, the proposed functional
form of $\tau$ is flexible enough to encompass both absorption models
and energy loss models. The 2 classes of models are distinguished by 
the value of the scaling exponent: a positive $\lambda \gneqq 0$ is
characteristic of absorption models, while a negative $\lambda \lesssim
0$ is characteristic of energy loss models. Thus, the exponent
$\lambda$ obtained in the proposed model-independent 
scaling analysis of experimental data can identify the leading
mechanism for hadron suppression in nDIS. 
%As discussed in
%Section~\ref{sec:scalinganalysis}, HERMES experimental data favor
%hadron absorption as opposed to energy loss. 

\section{Scaling of $\bm R_M$}
\label{sec:scalingtheory}

The idea that the hadron multiplicity ratio $R_M$ should scale with
the variable $\tau$ introduced in Eq.~\eqref{eq:scalingvar} 
is quite natural in the context of hadron absorption models
\cite{AMP03,AGMP05,Kopeliovich,Falteretal04,BG87}. 
In these models the struck quark neutralizes its color on a relative
short time scale. The ensuing color neutral state, called a
prehadron, later on collapses on the wave function of the observed
hadron. Hadron suppression is then mainly attributed to
prehadron-nucleons interactions, whose magnitude depends on the
in-medium prehadron path length, which depends solely on
the prehadron formation time $t_*$.

Estimates of the prehadron formation time can be obtained in the
framework of the Lund string model \cite{BG87,AMP03,AGMP05}, where the
prehadrons are identified with each of the fragments of the color
string.
%and their formation time with the time at which a
%quark-antiquark pair is created out of the vacuum and breaks the string.
Alternatively, in the pQCD inspired color dipole model for leading
hadron suppression of 
\refref{Kopeliovich}, the hardest gluon radiated off the struck quark 
splits into a quark-antiquark pair; the antiquark then recombines
with the struck quark into the leading prehadron. In both 
cases the prehadron formation time has a simple general form:
\begin{align}
  t_* & = g(z_h) (1-z_h) \frac{\nu}{\kappa} \ ,
 \label{eq:formtime}
\end{align}
where $g(z_h)\ra 0$ as $z_h\ra 0$, and
$\kappa$ is a constant that sets the time scale of hadronization. In
the Lund model $\kappa \approx 1$ GeV/fm is given by the string
tension; in the dipole model $\kappa = Q^2$. 
At HERMES, in both models, the prehadron formation time is $t_* \lesssim 
5$ fm, which is smaller than the nuclear radius. On the contrary,
hadrons are typically produced at the periphery or outside the target
nucleus and their absorption does not contribute much to $R_M$. 
The physical origin of $t_*$ is transparent. The factor $\nu$ can
be understood as a Lorentz boost factor. At large $z_h$ the
hadron carries away most of the struck quark energy. The color string
remainder has only an energy $(1-z_h)\nu$ left, so that it cannot
stretch farther off (in pQCD terms, the colored struck quark has a
little energy to radiate into gluons, hence it must neutralize its
color in a short time).
At small $z_h \ra 0$ the prehadron formation time should go to 0, 
as well, as explicitly shown in Lund model computations 
\cite{BG87,AMP03,AGMP05}. This follows from the fact that we are discussing
semi-inclusive hadron measurements. At small $z_h$ the observed 
hadron carries away a small fraction of the struck quark energy. The 
rest of the energy will most probably be used for the creation of other
low energy prehadrons, because the string fragmentation function is steeply
falling with $z_h$. On average, the observed prehadron will be
produced close to the interaction point.

Summarizing the above discussion, $R_M$ in absorption models 
should depend only on $t_*=t_*(z_h,\nu)$ and
not on $z_h$ and $\nu$ separately. A good approximation to $t_*$ is
the scaling variable $\tau$ of Eq.~\eqref{eq:scalingvar}, where the
scaling exponent $\lambda$ depends on the chosen absorption model.
A rough estimate of the scaling exponent gives $\lambda
\approx 1$. A more precise value can be obtained by fitting
\eqseqref{eq:RMscaling}-\eqref{eq:scalingvar} to the theory model
results \cite{AGMP05,Accardi05,Kopeliovich,Falteretal04} 
for $R_M$. The fit procedure, explained in detail in the next section, 
results in $0.5 \lesssim \lambda \lesssim 1.2$ for absorption
models. 

In energy loss models \cite{Wang,Arleo} the hadron formation time is
assumed to be much larger than the nuclear radius, and the
hadronization process is assumed to happen entirely outside the target
nucleus \footnote{In
  Ref.~\cite{Arleo}, the effect of stopping the gluon radiation after
  a finite prehadron formation time, but without any prehadron-medium
  interaction, has also been explored.}.
The quark travels through the nucleus and experiences multiple
scatterings and induced gluon bremsstrahlung. Hence, it starts
the hadronization process with a reduced energy $\nu-\epsilon$ where
$\epsilon$ is the energy of the radiated gluons 

In \refref{Arleo}, extended in \cite{Accardi05} to include finite
medium size corrections, the reduced quark energy at the time of
hadronization is translated into a shift of $z_h$ in the vacuum
fragmentation function $D$ \cite{WHS96}. The medium modified FF is then
computed as  
\begin{align}
  \tilde D_A(z_h) = \int_0^{(1-z_h)\nu} d\epsilon \, {\cal P}(\epsilon) \, 
    \frac{1}{1-\epsilon/\nu} D\Big(\frac{z_h}{1-\epsilon/\nu}\Big) \ ,
 \label{eq:modFF-BDMS}
\end{align}
where the dependence of the vacuum FF on the hard scale scale $Q^2$ of
the process is understood, 
and the quenching weight ${\cal P}(\epsilon)$ is the probability 
distribution of an energy loss $\epsilon$ computed
in the Baier-Dokshitzer-Mueller-Schiff formalism \cite{BDMS}. 
Note the upper limit of integration in \eqeqref{eq:modFF-BDMS} 
imposed by energy conservation. For the purpose of discussing the
scaling properties of $R_M$, we can work in the soft gluon
approximation, and neglect finite quark energy corrections, which
would introduce an 
additional $\nu$ dependence in the quenching weight \cite{Arleo02}.
If we further neglect energy loss fluctuations, we can 
approximate $R_M\approx \widetilde D_A(z_h)/D(z_h)$ and obtain
\begin{align}
  R_M \approx  \frac{1}{1-\vev{\epsilon}/\nu} 
    D \Big( \frac{z_h}{1-\vev{\epsilon}/\nu} \Big) 
    \, \big[ D(z_h) \big]^{-1} \ ,
 \label{eq:approxRM}
\end{align}
where the average energy loss $\vev{\epsilon} = \int_0^{(1-z_h)\nu}
d\epsilon\,\epsilon\,{\cal P}(\epsilon) / \int_0^{(1-z_h)\nu}
d\epsilon\,{\cal P}(\epsilon) = f[(1-z_h)\nu]$ is a
function of the energy $(1-z_h)\nu$ not carried away by the observed
hadron.
Next, we can approximate the FF using the parametrization of
\refref{KKP00} at $Q^2=2$ GeV$^2$: $D(z_h) = C z_h^\alpha
(1-z_h)^\beta$, where for pions $\alpha \approx -1$, $\beta \approx
1.5$ and the constant $C$ will cancel in the multiplicity ratio. 
Finally,
\begin{align}
  R_M \approx 
    \frac{1}{\Big( 1 - \ds\frac{1}{\nu} f[(1-z_h)\nu] \Big)^{\alpha+\beta+1}}  
    \left( 1 - \frac{f[(1-z_h)\nu]}{(1-z_h)\nu} \right)^\beta 
 \label{eq:scalingArleo}
\end{align}
which shows an approximate scaling with $(1-z_h)\nu$.

In \refref{Wang} the medium modifications of the fragmentation
functions are computed from twist-4 contributions to the leading order
cross-section, including diagrams with one elastic quark-nucleus 
scattering and one radiated gluon. Both the struck quark and the
radiated gluon are allowed to fragment according to vacuum FF. The
obtained modified FF, $\tilde D$, can be well approximated by the
numerator in \eqeqref{eq:approxRM} 
with $\epsilon/\nu = 0.6 \vev{z_g}$, where
$\vev{z_g}$ is the average fractional energy of the radiated gluon
\cite{Wang,GW01}:
\begin{align} 
 \vev{z_g} & = \int_0^{\mu^2}\frac{d\ell_T^2}{\ell_T^2} 
    \int_0^{1-z_h} dz_g \frac{\alpha_s}{2\pi} z_g \,
    \Delta\gamma_{q\rightarrow gq}(z_g,\ell_T^2) \nonumber \\
  & \approx \alpha_s^2(Q^2)\tilde C(Q^2) m_N R_A^2
    \frac{1}{\nu} f_g(1-z_h) \nonumber \\
  & \equiv \frac{k}{0.6} \frac{1}{\nu} f_g(1-z_h)\ .
\end{align}
Here, $\gamma_{q\rightarrow gq}$ is the quark-gluon splitting
function, $\tilde C(Q^2)$ is the strength of parton-parton correlations in
the nucleus, $m_N$ the nucleon mass, $R_A$ the nuclear radius, and $k$
is a shorthand for the quantities independent of $z_h$ and $\nu$. 
$f_g$ is s function of $1-z_h$ because of the upper limit of
integration on $z_g$ imposed by energy conservation. In the HERMES
regime, $f_g(1-z_h)\propto (1-z_h)^{0.4}$.
Approximating $R_M$ and the modified FF as before we have:
\begin{align}
  R_M \approx \frac{1}{\big( 1 - \frac{k}{\nu}f_g(1-z_h)
    \big)^{\alpha+\beta+1}} 
    \left( 1 - \frac{k \, f_g(1-z_h)}{(1-z_h)\nu} \right)^\beta \ .
 \label{eq:scalingWang}
\end{align}

From Eqs.~\eqref{eq:scalingArleo} and \eqref{eq:scalingWang} a scaling
of $R_M$ with $(1-z_h)$ is evident, which implies $\lambda=0$ in
\eqeqref{eq:scalingvar}. However, it is not immediate to see the
role played by $\nu$. To establish it, let's introduce an effective
scaling variable 
$%\begin{align}
  \tau' = C z_h^\lambda (1-z_h) \nu^\mu
$, %\end{align}
with $\mu$ an effective parameter describing the scaling of $R_M$ with
respect to $\nu$ in energy loss models. 
The value of $\mu$ can be determined by a fit of the full
computations in Refs.~\cite{Wang,Accardi05} as follows. 
For any given $\mu$, we fit
the theoretical $R_M=R_M(\tau')$ and determine
$\lambda=\lambda(\mu)$ by $\chi^2$ minimization as described in 
Section \ref{sec:fitprocedure}. A scaling of $R_M$ with $\tau'$
(i.e., $\chi^2/{\rm d.o.f.} < 1$) is found 
for $0.2 \lesssim \mu \lesssim 1.8$, with the 
best-fit $\lambda$ decreasing as $\mu$ increases. 
As also expected on
theoretical grounds, in this range of $\mu$ values 
one finds $\lambda_{\rm best} \lesssim 0$, 
which distinguishes it from the positive $\lambda$ expected in
absorption models.  

\begin{figure}[tb]
  \vspace*{0cm}
  \centerline{
  \includegraphics
    [width=0.85\linewidth,bb=18 144 540 718,clip=false]
    {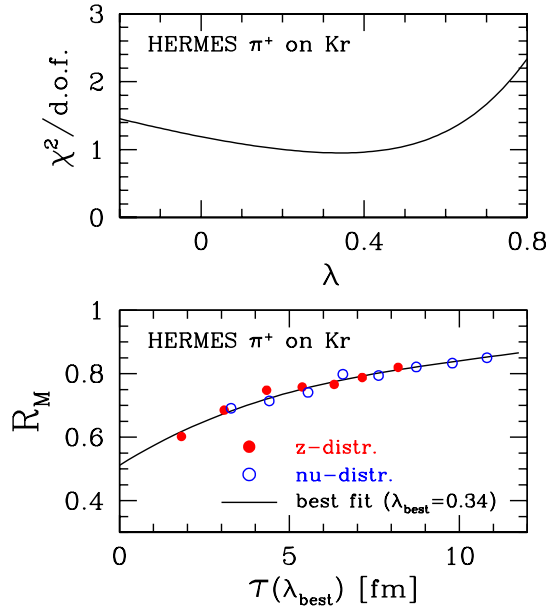}
  }
  \vspace*{0cm}
  \caption[]{
  An example of the fit procedure described in Section 3, applied to 
  $\pi^+$ production on a Kr target at HERMES \cite{HERMES2}. Upper panel:
  $\chi^2$ as a function of $\lambda$. Lower panel: $R_M(\tau)$ with
  $\tau$ computed at $\lambda_{\rm best}=0.34$. Experimental statistical
  errors are of the same size as the plotted points. 
  \label{fig:lambdafit}
 }
\end{figure}

In conclusion, for the sake of comparing energy loss models with
absorption models, where $\mu=1$ is theoretically justified, I will fix 
$%\begin{align}
  \mu=1
$ %\end{align}
and analyze experimental data and theory models 
in terms of the scaling variable $\tau$ proposed in
\eqeqref{eq:scalingvar}. 
%With many more data points than currently available from the HERMES
%experiment, one might try a 2 parameter fit in order to
%explore the correleations between $\lambda$ and $\mu$. 

Finally, a note on the limitations of the proposed scaling. 
In absorption models, the proposed
scaling might be broken by the dependence of the prehadron cross
section $\sigma_*$ on the photon virtuality $Q^2$ and the prehadron
energy $E_*\approx z_h\nu$, and by a possible
dependence of $t_*$ on $Q^2$ \cite{Kopeliovich}. At present, all these
effects cannot be calculated from first principles, and are to a good
extent model dependent. The $Q^2$ dependence
of $\sigma_*$ and $t_*$ is not a concern for the analysis of HERMES
data because the range of $\vev{Q^2}$ in their $z_h$- and $\nu$-bins
is rather small, implying a small scale breaking. 
In future experimental analyses it will be important
either to measure $R_M$ at fixed $Q^2$ or to ensure that $\vev{Q^2}$
stays approximately constant in all bins.
Since the prehadron must evolve into the observed hadron on a
relatively short timescale, one may expect that  $\sigma_* \propto
\sigma_h(E_h)$ when averaging $\sigma_*$ along the
prehadron path \cite{AMP03,AGMP05}. Since $\sigma_h$ has a mild
dependence on the hadron energy $E_h$ in the HERMES kinematic 
regime, only minor deviations from scaling are expected.
In Ref.~\cite{Kopeliovich}, $\sigma_*$ is computed in the
pQCD dipole model and is explicitly $E_h$ and $Q^2$ dependent. 
The good fit of $\lambda$ obtained in
Section~\ref{sec:scalinganalysis} for this model 
is an {\it a posteriori} indication
that scale breaking effects are small.
In energy loss models, scale breaking may arise due 
to fluctuations in energy loss, especially near the kinematic limit.
The relatively large error bars on $\lambda$ found in the
fits of Section~\ref{sec:scalinganalysis} show that their effect is
not fully negligible
\footnote{Since $\nu = O(10 {\rm\ GeV})$, scale breaking may also
  be caused by finite quark energy corrections \cite{Arleo}. However they have 
  not yet been
  implemented in the quenching weights \cite{BDMS} used in the
  computations of \cite{Accardi05} analyzed in this paper.}.  
However, this does not spoil the discriminative 
power of $\lambda$, which yields $\lambda \lesssim 0$ for energy loss
models but $\lambda \gneqq 0$ for absorption models.

\begin{figure}[tb]
  \vspace*{0cm}
  \centerline{
  \includegraphics
    [width=0.93\linewidth,bb=18 194 560 667,clip=true]
    {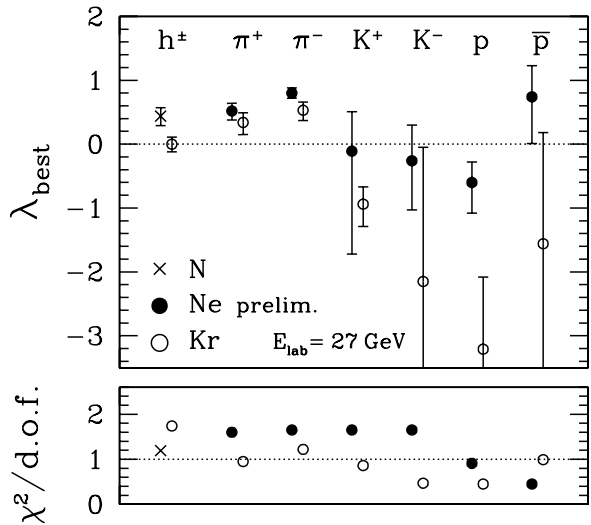}
  }
  \vspace*{0cm}
  \caption[]{
  The scaling exponent $\lambda_{\rm best}$
  extracted from HERMES data on charged and identified hadrons at
  $E_{lab}=27$ GeV \cite{HERMES1,HERMES2,HERMES3} (only 
  statistical errors included in the fit). Error bars correspond to 1 standard
  deviation. Bottom panel: $\chi^2$
  per degree of freedom.
  \label{fig:HERMESfit27}
 }
\end{figure}

\section{Fit procedure}
\label{sec:fitprocedure}

The HERMES experiment measures $R_M$ binned in $z_h$ and integrated
over $\nu$ and $Q^2$ (``$z_h$ distributions'') or binned in $\nu$ and
integrated over $z_h$ and $Q^2$ (``$\nu$ distributions''). 
The scaling of experimental 
data with respect to the variable $\tau$
defined in \eqeqref{eq:scalingvar} and the scaling exponent $\lambda$ 
can be determined by a fit to the data as follows.
\begin{enumerate}
\item[(1)]
  Fix $\lambda$.
\item[(2)]
  For each $z_h$ bin in $z_h$-distributions compute
  $\tau=\tau(z_h,\vev{\nu(z_h)})$ and $R_M(\tau)\equiv
  R_M(z_h)$, where $\vev{\nu(z_h)}$ is the average measured $\nu$ in
  the considered $z_h$-bin. 
  Likewise for each $\nu$ bin in $\nu$-distributions compute
  $\tau=\tau(\vev{z_h(\nu)},\nu)$ and $R_M(\tau)\equiv R_M(\nu)$.
\item[(3)]
  Fit a function $\phi(\tau)$ to the pairs $\{(\tau,R_M)\}$ obtained
  at step 2, and compute $\chi^2=\chi^2(\lambda)$. The choice of $\phi$
  is discussed below.
\item[(4)]
  Determine the best-fit exponent $\lambda_{\rm best}$ by
  minimization of $\chi^2(\lambda)$.
\item[(5)]
  If $\chi^2(\lambda_{\rm best}) \lesssim 1$ per degree
  of freedom, we say that
  the analyzed data set scales with respect to $\tau$
  and is characterized by a scaling exponent $\lambda_{\rm best}$.  
\end{enumerate}
An example of this procedure and the corresponding $R_M(\tau)$
computed at $\lambda=\lambda_{\rm best}$ is illustrated in
Fig.~\ref{fig:lambdafit}. 
The fit to theoretical computations is done in the same way as the
fit to  HERMES data, by
considering the computed $R_M$ at the central value of each $z_h$ and $\nu$
experimental bin. Theoretical errors are estimated as
6\% of $1-R_M$ for the models of
Refs.~\cite{AGMP05,Accardi05,Falteretal04,Wang}, which need to fit 1
parameter to $R_M$ data \cite{AGMP05}, and 10\% for the model of
Ref.~\cite{Kopeliovich}.

\begin{figure}[tb]
  \vspace*{0cm}
  \centerline{
  \includegraphics
    [width=0.83\linewidth,bb=18 330 560 662,clip=true]
    {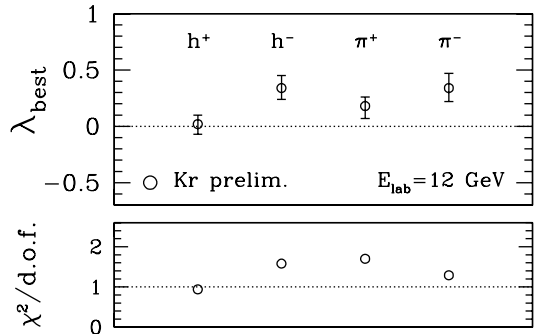}
  }
  \vspace*{0cm}
  \caption[]{
  The scaling exponent $\lambda_{\rm best}$ for HERMES data at $E_{\rm
  lab}$ =12 GeV on a Kr target \cite{vanderNat}. The N
  target results are not shown because of their very large error bars.
  Bottom panel: $\chi^2$ per degree of freedom.
  \label{fig:HERMESfit12}
 }
\end{figure}

The fit results discussed below have been obtained using  
as fit function $\phi(\tau)$ a polynomial of 4$^{\rm th}$ degree in
$\tau$. The results of the fit have been cross-checked by using a rational
function of second order constrained to tend to 1 as $\tau\ra\infty$,
which has 5 free parameters as the default polynomial. 
A second cross-check was obtained by additionally constraining
the rational function to have null derivative at $\tau = 0$ in order
to avoid singularities. 
%The advantage  
%of the polynomial function is that it gives a smooth
%$\chi^2(\lambda)$, while the fit with rational functions sometimes
%fails resulting in a non smooth $\chi^2(\lambda)$. 
%Sometimes, one finds a second minimum of $\chi^2(\lambda)$ 
%with comparable $\chi^2_{\rm best}$ but at much larges $\lambda \gg
%1$. This minimum is neglected, because it is unphysical in both absorption
%and energy loss models, and it is not confirmed by all 3 fit functions
%considered. 
Finally, unless otherwise explicitly stated, 
for this scaling analysis I considered only data
points satisfying the following cuts.
(i) $z_h>0.2$ ($\vev{z_h}>0.2$ for $\nu$-distributions), to avoid the
target fragmentation region and feed-down 
of hadrons from higher-$z_h$, for which the
conjectured scaling is not valid, and to avoid  large corrections due
to the detector geometric acceptance \cite{Falteretal04}. A cut at
$z_h>0.3$ or 0.4 might be preferable from this point of view, but
excessively reduces the available number of data points.
(ii) $z_h<0.9$, to avoid diffractive hadron production and
quasi-elastic lepton-nucleus scatterings.
(iii) $\nu>7$ GeV, for consistency between the analysis of the N,
Kr and Ne target data sets. For each target and hadron flavor 15
data points survive these cuts. I explicitly checked the stability of
$\lambda$ against small variations of the upper $z_h$ cut. 
The stability of $\lambda$ against variations of the lower $z_h$ cut 
is difficult to establish because of a rapidly shrinking number of
data points with increasing $(z_h)_{min}$. 

\begin{figure}[tb]
  \vspace*{0cm}
  \centerline{
  \includegraphics
    [width=\linewidth,bb=20 169 530 635,clip=true]
    {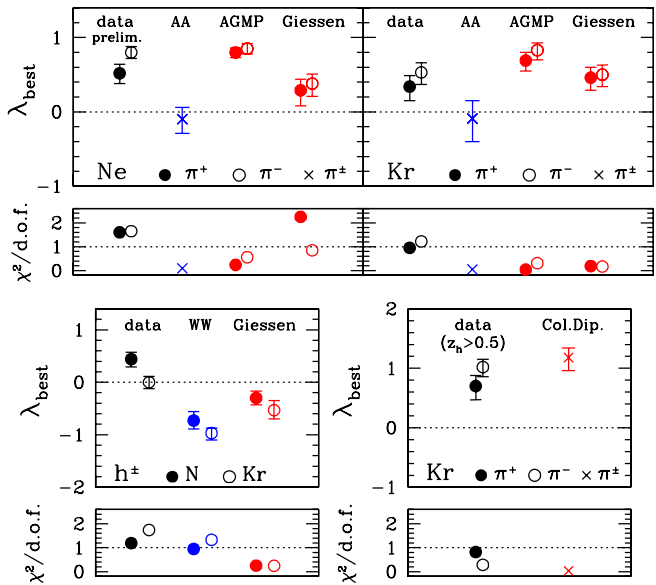}
  }
  \vspace*{0cm}
  \caption[]{
  Comparison of the scaling exponent for $\pi^\pm$ and $h^\pm$ 
  from HERMES data at $E_{\rm lab}=27$ GeV
  \cite{HERMES1,HERMES2,HERMES3} and from theory models. 
  Error bars correspond to 1 standard deviation.
  Energy loss models (blue points on-line): 
  AA \cite{Accardi05}, WW \cite{Wang}. Absorption
  models (red points): AGMP (pure absorption
  without $Q^2$-rescaling) \cite{AGMP05,Accardi05}, 
  Col.Dip. \cite{Kopeliovich}.  
  The Giessen model \cite{Falteretal04} embeds nuclear absorption in a
  full Monte Carlo simulation of the nDIS event. 
  Bottom panels: $\chi^2$ per degree of freedom. 
  \label{fig:datatheory}
 }
\end{figure}

\section{Results}
\label{sec:scalinganalysis}

The scaling exponents $\lambda_{\rm best}$ extracted from HERMES data
at $E_{lab}=27$ GeV \cite{HERMES1,HERMES2,HERMES3} and 12 GeV
\cite{vanderNat} for different hadron flavors produced on N, Ne and Kr targets 
are shown in Fig.~\ref{fig:HERMESfit27}. In all cases $\chi^2/{\rm
  d.o.f.} \lesssim 1.6$, which proves that $R_M$ scales with $\tau$.
The central result of this paper is that pion data 
exhibit a clear $\lambda_{\rm best} \gneqq 0$. As discussed in
Section~\ref{sec:scalingtheory}, this result contradicts the
assumption used in energy loss models that the quark is long-lived. 
In other words, it indicates the dominance of the prehadron
absorption mechanism as opposed to the energy loss
mechanism. This conclusion 
is confirmed by the comparison to theory models shown in
Fig.~\ref{fig:datatheory}. 
Therefore, the scaling variable $\tau$ can be interpreted
as a measure of the prehadron formation time. 
%Pions also show an indication of a small
%A-dependence of $\lambda_{\rm best}$, 
%which is unexpected in absorption models, and needs confirmation from
%the upcoming HERMES data on Xe targets.

Unidentified charged hadrons ($h^\pm$) have a positive 
$\lambda_{\rm best}$ on N
target, but $\lambda_{\rm best}\approx 0$ on Kr target. This apparently
contradictory result can be explained in terms of the proton 
contribution to the $h^\pm$ sample. Proton production shows an
anomalous enhancement of $R_M$ above 1 when $z_h \lesssim 0.4$, which cannot be
explained in terms of either parton energy loss or prehadron
absorption, and is not yet fully understood theoretically.
%It is in part due to feed-down from
%rescatterings of prehadrons produced at higher energy
%\cite{Falteretal04}, but other mechanisms are needed for a full
%understanding. 
The proton anomaly explains the negative value of its 
scaling exponent $\lambda_{\rm best}$, which in turn drives the
$h^\pm$ value of $\lambda_{\rm best}$ towards 0 for heavy
targets. Indeed, by cutting the $h^\pm$ 
sample on Kr target at $z_h > 0.5$ one obtains a reduced 
$\chi^2/{\rm d.o.f}=1.18$ and $\lambda_{\rm best}=0.34\pm0.13$,
compatible with the $\pi^\pm$ exponents and the nuclear absorption
mechanism. A further confirmation of the role of proton anomaly in
reducing the scaling exponent comes 
from preliminary data on Kr at $E_{lab} = 12$ GeV \cite{vanderNat},
which yield
$\lambda_{\rm best}=0.02\pm0.09$ for $h^+$ but
$\lambda_{\rm best}=0.34\pm0.11$ for $h^-$, see
Fig.~\ref{fig:HERMESfit12}.
From kaons and antiprotons data it is difficult to draw any conclusion
because of the large error bars.

\begin{figure}[tb]
  \vspace*{0cm}
  \centerline{
  \includegraphics
    [width=0.85\linewidth,bb=18 320 560 718,clip=true]
    {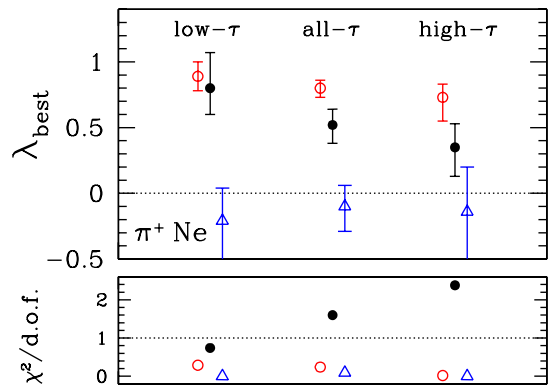}
  }
  \vspace*{0cm}
  \caption[]{
  Scaling exponent and $\chi^2$ per degree of freedom for   
  ``low-$\tau$'' and ``high-$\tau$'' data sets defined in
  Section~\ref{sec:scalinganalysis}, compared to the full data set for
  pion production on Ne targets, corresponding to ``medium-$\tau$''. 
  Black disks: $\pi^+$ from
  preliminary HERMES data \cite{HERMES3}. Red circles: $\pi^+$ from the
  AGMP absorption model \cite{AGMP05,Accardi05}. Blue triangles:
  $\pi^\pm$ from the energy loss model of Ref.~\cite{Arleo,Accardi05}.
  (Color on line).
  \label{fig:taucuts}
 }
\end{figure}

An interesting cross-check of the interpretation of $\tau$ as the
prehadron formation time can be obtained by dividing the full data set
in two subsets with low- and high-$\tau$. In the high-$\tau$ data set
the prehadron has a shorter in-medium path length, and the quark has a
longer in-medium quark path length, than in the low-$\tau$ data set.
Then, one expects a smaller contribution of prehadron absorption and a
larger contribution from partonic interactions and energy loss, hence 
a smaller $\lambda$ \footnote{This can be confirmed by stopping induced
  gluon radiation after a finite Lund prehadron formation time 
  in the energy loss model of Ref.~\cite{Arleo,Accardi05}. The fitted
  $\lambda\approx 0.4$ is intermediate between the pure energy loss 
  $\lambda\approx 0$ of \cite{Arleo,Accardi05} 
  and the pure absorption $\lambda\approx 0.8$ of  \cite{AGMP05,Accardi05}.}.
Since $\tau$
monotonically decreases with $z_h$ in HERMES $z_h$-distributions, and
monotonically increases with $\nu$ in  $\nu$-distributions, we can
define the 2 subsets by the following cuts on $z_h$ and $\nu$:
\begin{itemize}
\item
  low-$\tau$:
  $0.5 < z < 0.9$ and $7 {\rm\ GeV} < \nu < 13 {\rm\ GeV}$
\item
  high-$\tau$:
  $0.2 < z < 0.6$ and $13 {\rm\ GeV} < \nu $ ,
\end{itemize}
with 8 data points each, and a reasonable overlap of $z_h$- and
$\nu$-distributions. The partial overlap in the $z_h$-cuts is used to
improve the statistics of the 2 data sets.
%and reduce as much as possible the error bars on the corresponding
%$\lambda$. 
The full
data set defined in Section~\ref{sec:fitprocedure} is characterized by
an average $\tau$ intermediate between the 2 above data sets. The
fitted $\lambda$ for $\pi+$ production on Ne are plotted in
Fig.~\ref{fig:taucuts}. Preliminary HERMES data on a Ne target \cite{HERMES3} 
is compared with the absorption model of Ref.~\cite{AGMP05,Accardi05}
and the energy loss model of Ref.~\cite{Arleo,Accardi05}.
Though the error bars in the 2 subsets are relatively large, 
experimental data hint at a decrease of $\lambda$ with $\tau$, which
confirms the interpretation of $\tau$ as prehadron formation time. Its
modest slope indicates that induced partonic energy loss in cold
nuclear matter is rather weak, as also predicted in
Ref.~\cite{Kopeliovich}.

\section{Summary and conclusions}
\label{sec:conclusions}

In this work, I proposed a scaling analysis of hadron attenuation in
nDIS as a tool to investigate quark hadronization in cold nuclear matter,
and to distinguish parton energy loss from nuclear absorption
effects in experimental data.
The scaling properties of experimental data and theory computations of
the hadron attenuation ratio $R_M$ can be summarized by the value of
the exponent $\lambda$ in the scaling variable $\tau$ introduced in
\eqeqref{eq:scalingvar}. The exponent 
$\lambda$ is able to clearly distinguish models based on parton energy
loss ($\lambda \lesssim 0$) from models based on hadron absorption
($\lambda \gtrsim 0.5$). Experimental data on pion and 
charged hadron production have been shown to scale with $\tau$ and
exhibit $\lambda \gtrsim 0.4$, which is a clear indication
that the hadronization process starts on a time scale of the order of
a few Fermi, and that prehadronic nuclear absorption dominates hadron
quenching in nuclear DIS. The scaling variable $\tau$ can then be
interpreted as a measure of the formation time of the prehadron, the
color neutral precursor of the observed hadron. Note that the scaling
analysis cannot measure the absolute magnitude of the prehadron
formation length, only its dependence on $z_h$ and $\nu$.  
A more direct detection of in-medium hadronization, and a measurement
of the overall scale of the prehadron formation time, is possible by
looking at the hadron $p_T$-broadening, as proposed in
Ref.~\cite{Kopeliovich}. The scaling analysis described in this paper 
will be a useful cross-check of this measurement.
%An independent test of these conclusions, and a measure of the overall
%normalization of the prehadron  
%formation time $\tau$, can be obtained considering more exclusive
%observable, like the $z_h$-dependence of  hadron's $p_T$-broadening and
%Cronin effect \cite{Kopeliovich}. 
Establishing a scaling of the
prehadron formation time with inverse $Q^2$, as predicted, e.g., in
Ref.~\cite{Kopeliovich}, will further constrain the hadronization
mechanism. A dedicated experimental analysis is needed to 
improve the reach and precision of the scaling analysis 
presented in this paper.
%which is based on the published data points for $R_M$.  
%Upcoming HERMES data on Xe target and the high statistics
%measurements possible with the CLAS experiment \cite{CLAS} will
%allow to test in detail the dynamics and space-time evolution of
%hadronization.   
\\

%%%%%%%%%%%%%%%%%%%%%%%%%%%%%%%%%%%%%%%%
%%%%%%%%% ACKNOWLEDGMENTS %%%%%%%%%%%%%%
%%%%%%%%%%%%%%%%%%%%%%%%%%%%%%%%%%%%%%%%

%\vskip1cm
%\begin{acknowledgments}

%\vskip.5cm
{\bf Acknowledgments.} 
I am grateful to P.~Di~Nezza, V.~Muccifora and J.W.~Qiu
for valuable discussions and their support; to F.~Arleo, T.~Falter and
K.~Gallmeister for detailed discussions on their models; to I.~Vitev
for valuable comments; to INFN Frascati and INFN Milan for hospitality
during the preparation of this work.   
This work is partially funded by the US Department of Energy grant
DE-FG02-87ER40371. 

%\end{acknowledgments}

%%%%%%%%%%%%%%%%%%%%%%%%%%%%%%%%%%%%%%%%%%%%%%%%%%%%%
%%%%%%%%%%%%%%%%%%%%%  BIBLIOGRAFIA  %%%%%%%%%%%%%%%%
%%%%%%%%%%%%%%%%%%%%%%%%%%%%%%%%%%%%%%%%%%%%%%%%%%%%%

\end{document}